# Anderson transition of in-gap quasiparticles in a quasi-two-dimensional disordered superconductor

Hae-Ryong Park[1,2]*, Kyung-Yong Park[1]*, Kyoung-Min Kim[3]*, Kyung-Hwan Jin, Jun-Sung Kim[1,2], Han-Woong Yeom[1,2], Ki-Seok Kim[1,4]†, Jhinhwan Lee[2]†

[1]*Department of Physics, POSTECH, Pohang, Gyeongbuk 37673, Republic of Korea*

[2]*Center for Artificial Low Dimensional Electronic Systems,*
*Institute for Basic Science (IBS), Pohang 37673, Republic of Korea*

[3]*Center of Theoretical Physics of Complex Systems,*
*Institute for Basic Science (IBS) Daejeon 34126, Republic of Korea*

[4]*Asia Pacific Center for Theoretical Physics (APCTP), Pohang, Gyeongbuk 37673, Republic of Korea*

**The Anderson transition of Bogoliubov-de Gennes (BdG) quasiparticles in superconducting state has been studied theoretically for last three decades. However, its experimental proof is lacking. In particular, the relationship of the superconducting order-parameter fluctuations and the Anderson transition of BdG quasiparticles have not been well understood. Our study, based on scanning tunneling microscopy measurements, investigates how BdG quasiparticles become Anderson-localized and delocalized as a function of energy and applied magnetic field in a quasi-two-dimensional Fe-based superconductor with sufficient zero-bias BdG quasiparticles. The anomalous multifractal spectra based on the spatial distributions of the pairing gaps and the coherent peak heights suggest that superconducting fluctuations play a key role in the delocalization of in-gap BdG quasiparticles. Our real-space Hartree-Fock-BCS-Anderson simulations and renormalization group analysis with pairing fluctuations support quasiparticle localization and suggest that enhanced pairing fluctuations lead to delocalization of BdG quasiparticles and 'weak localization' of phase-fluctuating Cooper pairs in quasi-two-dimensional disordered superconductors. The present study proposes that the 10-fold way classification scheme has to be generalized to take order-parameter fluctuations in actual quantum matter. Also, it shed light on how ac energy loss due to quasiparticles at Fermi level can be controlled in a quasi-2d superconductor with sufficient pairing fluctuation.**

Researches on the superconductor-to-insulator transitions driven by disorders have advanced our understanding of superconductivity in terms of the quantum coherence in Cooper pair and Bogoliubov-de Gennes (BdG) quasiparticle dynamics[1-11]. Two pathways have been proposed for these quantum phase transitions: phase-disorder-driven[12-22] and amplitude-fluctuation-governed[23,24]. The phase-disorder mechanism suggests Anderson localization of preformed Cooper pairs, leading to a pseudogap and lack of coherent peaks in the quasiparticle spectrum of the insulating state. On the other hand, the amplitude fluctuation case predicts Anderson localization of BdG quasiparticles, leading to the renormalization of Coulomb interaction to break Cooper pairs and the gap closure. However, strong inhomogeneity can mix these two pathways.

While the former has been extensively investigated both theoretically and experimentally, the Anderson transition of BdG quasiparticles inside the superconducting phase has been studied only theoretically[25-27]. As far as we know, the experimental proof of Anderson transition of BdG quasiparticles inside the superconducting phase has never been reported. In terms of experiments, the low-energy BdG quasiparticle states typically are not sufficiently generated with disorder to access the signature of localization clearly. Theoretically speaking, while its microscopic mechanism is elusive, Anderson transitions of BdG quasiparticles serve as a central building block in the ten-fold way classification scheme of topological matter[28,29], referred to as the Altland-Zirnbauer symmetry class[30]. Essential simplification in the ten-fold way classification scheme is that the superconducting order parameter remains to be fixed in the Anderson transition, where the dynamics of BdG quasiparticles is described by a non-interacting tight-binding lattice model. In this case the delocalization of in-gap BdG quasiparticles is allowed only at the zero energy, where other in-gap quasiparticle states are all localized, in two dimensions. The validity of this simplified model has not been tested yet.

The primary objective of this study is to investigate the Anderson transition of low-energy BdG quasiparticles in a disordered quasi-2d superconductor and its interplay with the superconducting pairing fluctuations. The experimental limitations in observing the localization of BdG quasiparticles is overcome by quasi-2d Fe-based superconductor with large spacings between adjacent superconducting monolayers and, above all, with significant pair breaking perturbations by the neighboring layers of magnetic V atoms leading to large zero-bias BdG quasiparticle density of states[2,31-33]. This system serves as a platform to directly observe the Anderson transition of BdG quasiparticles by spectroscopic imaging with scanning tunneling microscopy.

In this study, we propose the delocalization phenomenon starting from the energy close to the superconducting gap and approaching the Fermi level at stronger pairing fluctuations, which highlights the role of Cooper-pair fluctuations in the Anderson transition of in-gap BdG quasiparticles. Resorting to scanning tunneling microscopic (STM) measurements, we first confirm the abundance of zero-bias quasiparticle LDOS and the existence of static fluctuations of near-zero-bias quasiparticle LDOS with self-similarity up to a length scale sufficiently larger than the superconducting coherence length. We also obtain various statistical information from the distributions of the zero-bias quasiparticle LDOS ($\rho$), the local pairing gap ($\Delta$) and the coherent-peak height ($Z_\Delta$), and show how they change with applied magnetic field and investigate the localization properties of the quasiparticles and Cooper pairs. We also introduce novel multifractal spectra of $f_\Delta(\alpha; H)$ and $f_{Z_\Delta}(\alpha; H)$ given by multiple moments of the local pairing gaps and the coherence peak heights in addition to the conventional multifractal spectrum $f_\rho(\alpha; H)$ of the quasiparticle LDOS[34-36]. These two 'anomalous' multifractal spectra

$f_\Delta(\alpha; H)$ and $f_{Z_\Delta}(\alpha; H)$ together with the conventional multifractal spectrum $f_\rho(\alpha; H)$ suggest how the in-gap BdG quasiparticles become delocalized with applied magnetic field, shown in the schematic phase diagram Fig. 1a and the renormalization group (RG) based phase diagram Fig. 1b, which will be discussed below in detail. We claim that this study proposes a novel mechanism for delocalization of quasiparticle excitations in quasi-two dimensions beyond the two well-known mechanisms of the ten-fold way classification without electron correlations[28,29,37] or the appearance of strong ferromagnetic fluctuations[38,39].

To verify the magnetic-field driven delocalization of the in-gap BdG quasiparticles, we have taken local spectroscopic maps of the disordered quasi-2D Fe-based superconductor $Sr_2VO_3FeAs$ at 0 T and 7 T. See the STM measurement in Fig. 1c. These maps are taken at 4.8 K well below the superconducting $T_c$ of ~ 30 K, and the maximum magnetic field 7 T is well below $H_{c2}$ (>> 10 T) at 4.8 K[33,40,41]. Due to the significant separation of the adjacent FeAs layers by insulating $Sr_2VO_3$ bi-layers (Fig. 1c), as evidenced by the conductance in c direction 1/100 times smaller than those in the ab plane[33,41], the system could be a model quasi-2D FeAs superconductor to investigate the energy dependent localization of BdG quasiparticles, with the O vacancy defects in the $VO_2$ layers playing the role of potential and magnetic scattering centers for both the quasiparticle interference and the Anderson localization of quasiparticles. The Fermi-level proximity of the α band minimum and the tunneling matrix effect of the capping $Sr_2VO_3$ layer contribute to the overall asymmetry of the tunneling spectra (Fig. 1d). It is known that the bands of the capping $Sr_2VO_3$ layers are gapped and separated from the Fermi level at least by 200 meV and hence do not contribute to the zero-energy LDOS[41]. However, the magnetic $VO_2$ layer with Neel order near the FeAs layers may contribute to the zero-bias quasiparticles by time-reversal symmetry breaking[33,42].

The topograph simultaneously taken with the zero-field LDOS map is shown in Ext. Data Fig. 2a and the strongest quasiparticle scattering centers extracted from the -20 meV LDOS map of Ext. Data Fig. 2g are identified as O vacancies in neighboring $VO_2$ layers[43] and are overlaid as red dots in all real-space maps throughout this Report. The zero-bias quasiparticle LDOS map at 0 T (Fig. 1e) shows strong long-range static fluctuations in the vicinity of the defects (Ext. Data Fig. 2h) which disappears in the LDOS map taken at 7 T (Fig. 1f). The energy-resolved autocorrelation plots of the quasiparticle LDOS (Fig. 1g-h) shows that they are restricted to a narrow energy range $|E|$ < 3 meV symmetrically centered at the Fermi level. The QPI modulations due to the intra-band scattering of electrons in Γ-centered electron pocket[41,43] are symmetrically formed around each defect (Ext. Data Fig. 2) with a strong energy-dependence of the wavelength and a much weaker magnetic-field-dependence.

Fitting the zero-bias autocorrelations of the quasiparticle LDOS to $Ae^{-r/\xi} + Be^{-\kappa r}\cos(2k_F r - \phi)$, which is a combination of the exponential decay (∝ $A$:Anderson localization-dependent long-range correlation and localization-independent short-range residual electronic correlation) and the sinusoidal exponential decay (∝ $B$:quasiparticle scattering interference), shows strong field-dependence of the fitted values of $\xi$ (6.67 nm at 0 T vs. 1.99 nm at 7 T) and we identify the Anderson localization length scale as $\xi_{AL} \simeq \xi_{0T} \simeq 6.67$ nm and localization-independent short-range residual electronic correlation as $\xi_{7T} \simeq 1.99$ nm. Striking evidence for the existence of quasi-localized or critical quasiparticles at zero magnetic field can be given by the scale-dependent multifractal spectrum analysis of the zero-bias quasiparticle LDOS (Figs. 1j and 1k). Varying the box size for the multifractal spectrum whose formulation is described in SI section 1.3, the multifractal spectrum of the zero-bias quasiparticle LDOS remains to be self-similar (critical) within the Anderson localization length scale $\xi_{AL}$ at 0 T as shown in Fig. 1j and Fig. 2t.

The significantly large autocorrelation length scale $\xi_{AL}$ of the zero-bias LDOS at zero field compared with the 7 T case, its confinement within a narrow energy range centered at the Fermi level, and its disappearance by application of the magnetic field (Fig. 1h) indicate that quasi-localized or critical in-gap BdG quasiparticles at zero magnetic field become delocalized by applied magnetic field. Below, we discuss how this delocalization phenomenon is correlated with strong fluctuations of local pairing gaps.

First of all, it is useful to compare the quasiparticle localization length $\xi_{AL}$ with the Cooper-pair coherence length $\xi_\Delta$, given by the autocorrelation analysis of the zero-bias LDOS ($\rho$) and that of the pairing amplitude ($\Delta$), respectively. (See SI section 1.1 and Ext. Data Fig. 1b for the fitting procedure.) It turns out that $\xi_{AL} \approx 6.67$ nm is longer than $\xi_\Delta \approx 1.39$ nm at zero magnetic field. Although both length scales are reduced at 7 T, the inequality $\xi_\rho > \xi_\Delta$ still holds. We suspect that the inequality of $\xi_{AL} > \xi_\Delta$ may serve as a condition for the superconducting state to coexist with Anderson localization of BdG quasiparticles. The underlying physical picture is that the BdG quasiparticles remain coherent within the Cooper-pair coherence length scale, and such 'coherent' quasiparticle excitations allow global coherence of Cooper pairs, regarded to be 'BCS superconductivity'. This scenario can be verified by confirming $\xi_{AL} \approx \xi_\Delta$ in the vicinity of the superconductor-insulator transition.

To reveal the mutual correlation between the zero-bias quasiparticle LDOS and the local pairing gap, we extract out the local pairing-gap map and the coherence peak-height map as shown in Figs. 2a-f and calculate the cross-correlation of every map with respect to the zero-bias quasiparticle LDOS map (Fig. 2a) as shown in the insets of Figs. 2a-f. The coherence peak-height map has been discussed to play the role of the local Cooper-pair density map[11,22,44]. Excluding the QPI modulation length scale visible in Figs. 2a-f as concentric rings, there seems to be two other length scales in the cross-correlation of the zero-bias quasiparticle LDOS map and the local pairing-gap map (inset of Fig. 2b), where they are negatively correlated in short range (< 2 nm) due to the local anticorrelation of the zero-bias LDOS and the superconducting gap as predicted by the Dynes formula indicating local pair breaking effects, and positively correlated in long range (>2 nm) which will be discussed in detail below. On the other hand, the coherence peak-height map shows negative correlation with the zero-bias LDOS map in all length scales (inset of Fig. 2c) due to the much stronger spectral weight conservation shared between the zero-bias LDOS and the gap-energy LDOS.

To understand the positive long-range correlation between the zero-bias LDOS map and the local pairing-gap map, we recall the self-consistent gap equation in strongly disordered superconductors[18,21], given by $\Delta(\vec{r}) = \frac{g_{eff}}{2} \int d\epsilon \frac{\Delta(\epsilon)}{\sqrt{\Delta^2(\epsilon)+\epsilon^2}} |\psi(\epsilon,\vec{r})|^2 \tanh\frac{\sqrt{\Delta^2(\epsilon)+\epsilon^2}}{2T}$, where the energy-dependent gap function is $\Delta(\epsilon) = \int d^d\vec{r}\, \Delta(\vec{r}) |\psi(\epsilon,\vec{r})|^2$. Although $\psi(\epsilon,\vec{r})$ should be an eigenfunction in the normal state, we replace $|\psi(\epsilon,\vec{r})|^2$ with the LDOS measured in the superconducting phase. These two equations result in $\Delta(\epsilon) = \frac{g_{eff}}{2} \int d\epsilon'\, I(\epsilon,\epsilon') \frac{\Delta(\epsilon')}{\sqrt{\Delta^2(\epsilon')+\epsilon'^2}} \tanh\frac{\sqrt{\Delta^2(\epsilon')+\epsilon'^2}}{2T}$, where $I(\epsilon,\epsilon') = \int d^d\vec{r}\, |\psi(\epsilon,\vec{r})|^2 |\psi(\epsilon',\vec{r})|^2$ is the autocorrelation function. Using the LDOS in the superconducting state at 0 T, we obtain $I(\epsilon,\epsilon')$ in Fig. 2m, where

the bright spot at $(\epsilon, \epsilon') = (0,0)$ indicates Anderson localization at zero magnetic field. As a result, we find the energy-dependent gap function $\Delta(\epsilon)$ ($\Delta_{th}(\epsilon)$ in Fig. 2n) and convert it to its real space configuration $\Delta(\vec{r})$ ($\Delta_{th}(\vec{r})$ in Fig. 2o). Comparing Fig. 2o with Figs. 2b, we observe clear signature of the positive long-range correlation between the zero-bias LDOS and the pairing gap reproduced, which confirms that increased zero-bias LDOS due to Anderson localization is responsible for the increased pairing gap on the large scale of $\xi_{AL}$. Repeating this analysis at finite ranges of temperature $T$ and the BCS coupling constant $g_{eff}$ (Fig. 2p), we obtain critical temperatures as a function of the BCS coupling constant, which gives reasonable values ~ 30K at $g_{eff}$~0.6.

In a typical quasi-2d superconductor with negligible zero-bias LDOS, the applied magnetic field will reveal magnetic vortices as locally finite zero-bias LDOS. However, in the case of quasi-2d superconductor with strong pairing fluctuations and finite zero-field zero-bias LDOS, the major effect of magnetic field will be to delocalize the Anderson-localized quasiparticles and make the LDOS distribution more uniform. Also, the LDOS distribution will be generally upshifted due to partial closure of the pairing gap (Figs. 2g and 2j). The delocalization of the quasiparticles at 7 T is manifested by the narrowing of the multifractal spectrum at $L = 1$ (Fig. 2j) and the significant narrowing by half at $L \geq 16$ (9.376 nm) (Fig. 1k). The latter is free from the residual QPI-induced LDOS variance since the averaging box size $L$ is larger than the QPI modulation length scale $\frac{\pi}{k_F} \simeq 6.7$ nm and therefore closer to the multifractal spectrum for the ideal delocalization.

With the LDOS distribution being more uniform and pair breaking starting to take place in applied magnetic field, the superconducting gap distribution will be partially closed starting from the band with smallest pairing gap which may occur quasi-randomly at a length scale much shorter than the zero-field Anderson localization length scale. This will cause splitting of the pairing gap distribution into two parts, one peaked at finite values with similar distribution width compared with the zero-field case and another part newly peaked at zero-bias (Fig. 2h). This pair breaking scenario will lead to increased short-range fluctuations of measured pairing gap at high applied magnetic field manifested as widening of the multifractal spectrum of the pairing gap (Fig. 2k). On the other hand, at a length scale larger than $\xi_{AL}$, the positive long-range correlation of the pairing gap with the zero-bias LDOS discussed above (inset of Fig. 2b) will lead to decreased long-range fluctuations of the pairing gap at high applied magnetic field, following the behavior of the zero-bias LDOS. This will lead to the cross-over of the magnetic-field-dependence of the multifractal spectrum width at a length scale near $\xi_{AL}$ (Figs. 2q-2s).

The pair breaking by applied magnetic field will also strongly down-shift the distribution of the coherence peak height (Fig. 2i). It is interesting to see that the coherence peak height whose direct distribution width decreases with applied field (Fig. 2i) will have increased normalized distribution width leading to increased width of their multifractal spectra in applied magnetic field (Fig. 2l).

To understand the opposite multifractal behaviors (Figs. 2j and 2k) of the in-gap BdG quasiparticles and the local superconducting order parameter with minimal set of assumptions, we turn to the real-space Hartree-Fock-BCS-Anderson simulation[45,46] for two-dimensional disordered superconductors with s-wave pairing symmetry. We introduce the following BCS Hamiltonian defined on a two-dimensional square lattice,

$$H = -\sum_{\langle i,j \rangle \sigma}[t_{ij} + (\epsilon_i - \mu)\delta_{ij}]c_{i\sigma}^\dagger c_{j\sigma} + \sum_i (\Delta_i c_{i\uparrow}^\dagger c_{i\downarrow}^\dagger + \text{H.c.}),  \tag{1}$$

where $\Delta_i = U\langle c_{i\downarrow} c_{i\uparrow}\rangle$ is the superconducting order parameter and $\epsilon_i$ is an on-site random potential uniformly given in $(-V, V)$. Here, we consider the strong disorder regime of $V = 4$, where all single-particle states are localized in the normal state. We also introduce an effective repulsive interaction $U_r$, which reduces the effective random potential as $\epsilon_i \to \epsilon_i + \frac{U_r}{2}\langle n_i \rangle_0$ and enhances the density of electrons around the zero energy. See the SI section 2 for more details. $\mu$ is the chemical potential to adjust the occupation number. $t_{ij}$ is an effective hopping parameter between nearest neighbor sites. Considering an external uniform magnetic field applied in z-axis, this hopping parameter is modified as

$$t_{ij} c_{i\sigma}^\dagger c_{j\sigma} + h.c. = -t[c_{i+1,j}^\dagger c_{i,j} e^{i\Phi_x} + h.c. + c_{i,j+1}^\dagger c_{i,j} e^{i\Phi_y} + h.c.]. \quad (2)$$

The phase factors are given by $\Phi_x = \left(\frac{e}{\hbar}\right)\int_{x_i}^{x_j} A_x(y) dx$ and $\Phi_y = \left(\frac{e}{\hbar}\right)\int_{y_i}^{y_j} A_y(x) dy$. Here, we consider the symmetric gauge $\boldsymbol{A} = (-y\phi_0 p/q, x\phi_0 p/q, 0)$ and the magnetic flux per unit cell is $\oint \boldsymbol{A} \cdot d\boldsymbol{s} = Ba^2 = 2\phi_0 p/q$ with the flux quantum $\phi_0 = h/e$ and coprime integers $p$ and $q$ (See the SI section 2.3 and Ext. Data Fig. 7).

We find that the distribution of the local paring amplitude is Gaussian-like (Fig. 3h) without external magnetic field, where the mean value of $\langle \Delta \rangle$ is smaller than that of the clean superconducting system. Applying external magnetic field, the Gaussian distribution of Cooper pairs becomes broader and distorted. In particular, there appears finite population of the zero local pairing amplitude. Furthermore, the sign change of the local pairing amplitude, i.e., the $\pi$ phase shift of the Cooper pair is observed in the Cooper-pair distribution function. See Fig. 3h. The local-pairing multifractal spectrum becomes broader (Fig. 3k), which indicates existence of their strong fluctuations driven by external magnetic field. On the other hand, the LDOS of the in-gap quasiparticles are enhanced and become more homogeneous (Fig. 3g). The density distribution function is narrower and shifted to larger mean values of the LDOS. The multifractal spectrum shows weaker multifractality by the magnetic field (Fig. 3j). The probability density function of a low-energy eigenfunction (Fig. 3i) and the corresponding multifractal spectrum (Fig. 3l) show qualitatively similar features with those of the LDOS. All these behaviors in the simulation with minimal assumptions are consistent with our STM measurements, which indicates that magnetic-field driven 'localization' of Cooper pairs occurs at the same time with delocalization of in-gap BdG quasiparticles.

The magnetic-field driven delocalization of the in-gap BdG quasiparticles is verified by the scaling behavior of the critical exponent $\alpha_2$ of the inverse participation ratio (IPR)[37,47]. The scaling behavior negatively correlated with enhancement of the system size indicates that the in-gap BdG quasiparticles are Anderson-localized without external magnetic field (Fig. 3m). On the other hand, we find that the IPR critical exponent increases as enlarging the system size, driven by external magnetic field, which confirms metallicity of the in-gap BdG quasiparticles (Fig. 3n). In Fig. 3m, we identify the zero-field mobility edge of the in-gap BdG quasiparticle state inside the superconducting gap based on the scale invariance of the IPR exponent. We point out that the mobility edge of the in-gap BdG quasiparticle state is quite close to the coherent peak position. We suspect that the superconductor-insulator transition would occur when the mobility edge crosses the coherent peak position. It is not clear how $\xi_\Delta \approx \xi_{AL}$ discussed before can be consistent with this condition.

We further justify our real-space Hartree-Fock-BCS-Anderson simulation, investigating the energy dependence of the multifractal scaling exponent $\alpha_0$ (Figs. 3o-p) which quantifies the uniformity of measures and the strength of multifractality[1,2]. Both simulation and experimental results explain that the closer to the zero energy, the weaker the uniformity and the stronger the multifractality get.

To reveal the role of spatial fluctuations of pairing fields in delocalization of in-gap BdG quasiparticle states, we perform a renormalization group (RG) analysis[48]. Considering the fact that the superconductor belongs to the class C in the tenfold way topological classification[28,29], we write down a Lagrangian density of our minimal model as

$$L = \Psi^\dagger(-i\omega + vk\sigma_3 + \Delta_0\sigma_1)\Psi + \Psi^\dagger\Delta_1|k|^\alpha\sigma_1\Psi + \sum_{a=1}^{3} V_a\Psi^\dagger\sigma_a\Psi, \tag{3}$$

where $\Psi = (\psi_\uparrow, \psi_\downarrow^\dagger)$ is a Nambu spinor field that describes BdG quasiparticle states. Here, $\omega$ is an energy variable, $k$ is a radial momentum expanded about the Fermi momentum, $v$ is the Fermi velocity, $\Delta_0$ is a constant pairing field, and $\sigma_a$ are the Pauli matrices. $\Delta_1|k|^\alpha$ is a non-uniform pairing field introduced phenomenologically to describe disorder-induced in-gap states ($\alpha = 0.5$ is taken without loss of generality; See Fig. 4a). $V_1$, $V_2$ and $V_3$ stand for three disorder scattering terms, which originate from random pairing fluctuations in amplitude and phase, and a random potential, respectively. We assume a Gaussian white-noise distribution for each $V_a$, i.e, $\langle V_a(r)V_a(r')\rangle = \delta(r-r')\Gamma_a$, where the disorder parameter $\Gamma_a$ characterizes the variance of such a distribution. Therefore $\Gamma_1$ and $\Gamma_2$ are the measures of the pairing fluctuations increasing with the applied magnetic field and $\Gamma_3$ is a measure of the disorder strength. (See SI section 3 for more details about the construction of this model.)

Taking into account quantum corrections in the one-loop order, we obtain RG flow equations, which are given in an approximate form as (See SI section 3 for the full equations and computation details),

$$\frac{d\Delta_0}{dl} = \Delta_0[1 + A(-\Gamma_1 + \Gamma_2)], \tag{4}$$

$$\frac{d\Delta_1}{dl} = (1-\alpha)\Delta_1, \tag{5}$$

$$\frac{d\Gamma_1}{dl} = \Gamma_1 + B(\Gamma_1^2 + \Gamma_2^2), \tag{6}$$

$$\frac{d\Gamma_2}{dl} = \Gamma_2 + 2B\Gamma_1\Gamma_2, \tag{7}$$

$$\frac{d\Gamma_3}{dl} = \Gamma_3 - 2C\Gamma_1\Gamma_2, \tag{8}$$

where $l \to \infty$ in the zero-temperature limit and the coefficients $A, B, C$ are given by

$$A = \frac{1}{\omega^2+(\Delta_0+\Delta_1)^2}, \tag{9}$$

$$B = \frac{(\Delta_0+\Delta_1)^2}{[\omega^2+(\Delta_0+\Delta_1)^2]^2}, \tag{10}$$

$$C = \frac{\omega^2}{[\omega^2+(\Delta_0+\Delta_1)^2]^2}. \tag{11}$$

There exist two stable fixed-point (FP) solutions for these equations (See Fig. 4b):

$$\text{FP1}: \left(\frac{\Gamma_1}{\omega}, \frac{\Gamma_2}{\omega}, \frac{\Gamma_3}{\omega}\right) \to (\infty, \infty, \infty), \tag{12}$$

$$\text{FP2}: \left(\frac{\Gamma_1}{\omega}, \frac{\Gamma_2}{\omega}, \frac{\Gamma_3}{\omega}\right) \to (\infty, \infty, 0), \tag{13}$$

where FP1 (FP2) indicates that an in-gap BdG quasiparticle state for given $\omega$ is localized (delocalized). If $\Gamma_1$ and $\Gamma_2$ are so weak that FP1 is stabilized for all $\omega$, the system belongs to the conventional disordered superconducting phase[1-11,31,32] where all in-gap states are localized. On the other hand, if $\Gamma_1$ and $\Gamma_2$ are strong enough that FP2 is stabilized for certain $\omega$, the system belongs to a novel anomalous metallic phase with superconductivity where delocalized in-gap states are developed. In the latter case, the delocalized in-gap BdG quasiparticles may constitute metallic charge carriers in the superconducting gap, which could be reminiscent of the failed superconductor[15]. A quantum phase transition between two phases occurs when the disorder parameters are controlled (Fig. 4c).

Our detailed RG analysis taking the energy dependence demonstrates the delocalization nature of the quantum phase transition (See SI section 3 for the computation details). The energy dependent phase diagram in Fig. 4c explicitly shows the emergence of delocalized in-gap states as $\Gamma_1$ (the strength of superconducting fluctuations) increases. When $\Gamma_1$ is small, only localized in-gap states exist in an energy window (green region). When $\Gamma_1$ becomes larger, however, delocalized in-gap states also appear in a higher energy range (red region). A mobility edge (dashed line) appears in the latter case, which is comparable to that found in the numerical simulation (See Fig. 4). The phase diagram in Fig. 4e also exhibits the sharp transition of the energy range for delocalized in-gap states across the phase boundary in the disorder parameter space. The RG analysis suggests that our samples may undergo this delocalization transition, which is unveiled by the multifractality spectrum analysis in Fig. 2 (Exps.) and Fig. 3 (Sims.). When the magnetic field is absent, the pairing fluctuations are weak (small $\Gamma_1$ and $\Gamma_2$ in the effective field theory) and in the presence of magnetic field, the pairing fluctuations become strong (large $\Gamma_1$ and $\Gamma_2$). The RG analysis suggests that the former and latter belong to the conventional disordered superconducting phase and the anomalous metallic phase with superconductivity, respectively.

In summary, our STM measurements and real-space mean-field simulations confirmed that the quasiparticles are localized with self-similar fractal LDOS structure and are delocalized with applied magnetic field which enhances the Cooper-pair fluctuations rather strongly, evidenced by the Cooper-pair multifractal spectrum and the coherent-peak height-distribution function. Our disorder RG analysis suggests the existence of a fixed point characterized by relatively large variance of Cooper pairs, not controllable by the perturbative RG method. These results suggest an anomalous insulator-metal transition of in-gap BdG quasiparticles in a quasi-two-dimensional Fe-based superconductor, offering a novel mechanism for Anderson transitions in quasi-two dimensions beyond the ten-fold way classification.

We believe that the metal-insulator transition of in-gap BdG quasiparticles is not limited to the superconducting state only, but a ubiquitous phenomenon wherever the corresponding order-parameter field for spontaneous symmetry breaking is strongly fluctuating near the metal-insulator transition, and is expected to show the multifractal spectrum of the collective mode. Although we showed that the quasiparticle states can be delocalized by the fluctuations of order parameter fields, the fundamental mechanism is not fully clarified yet. It is natural to expect a two-parameter scaling theory for the Anderson transition based on the quasiparticle localization length and the order-parameter correlation length. STM measurements could reveal the interplay between the symmetry-breaking field and the corresponding quasiparticle state by the multifractal analysis of the collective spectrum.

**Data availability**

The source data are available from the corresponding authors upon reasonable request.

**Code availability**

The custom code and mathematical algorithms that support the findings of this study are available from the corresponding authors upon reasonable request.

**Acknowledgment**

We acknowledge useful experimental assistance by Seokhwan Choi and Jong-Mok Ok. The work was supported by the Institute for Basic Science (Grant No. IBS-R014-D1) and by National Research Foundation of Korea (Grant No. 2016R1D1A1B01016186, No. 2013M3C1A3064455, No. NRF-2021R1A2C1006453, and No. NRF-2021R1A4A3029839).

**Author contributions**

J.L. performed the STM experiment. H.R.P. and J.L. performed the data analysis. K.Y.P. performed the Hartree-Fock calculation, and K.M.K. performed the RG analysis. These theoretical analyses were under the supervision of K.S.K. J.S.K. synthesized and provided the bulk crystal. H.R.P., K.Y.P., K.M.K, H.W.Y., J.L., and K.S.K. co-wrote the manuscript. All authors discussed the results and commented on the paper.

**Competing financial interests**

The authors declare no competing financial interests.

**Corresponding authors**

Correspondence and requests for materials should be addressed to K.S.K. and J.L.

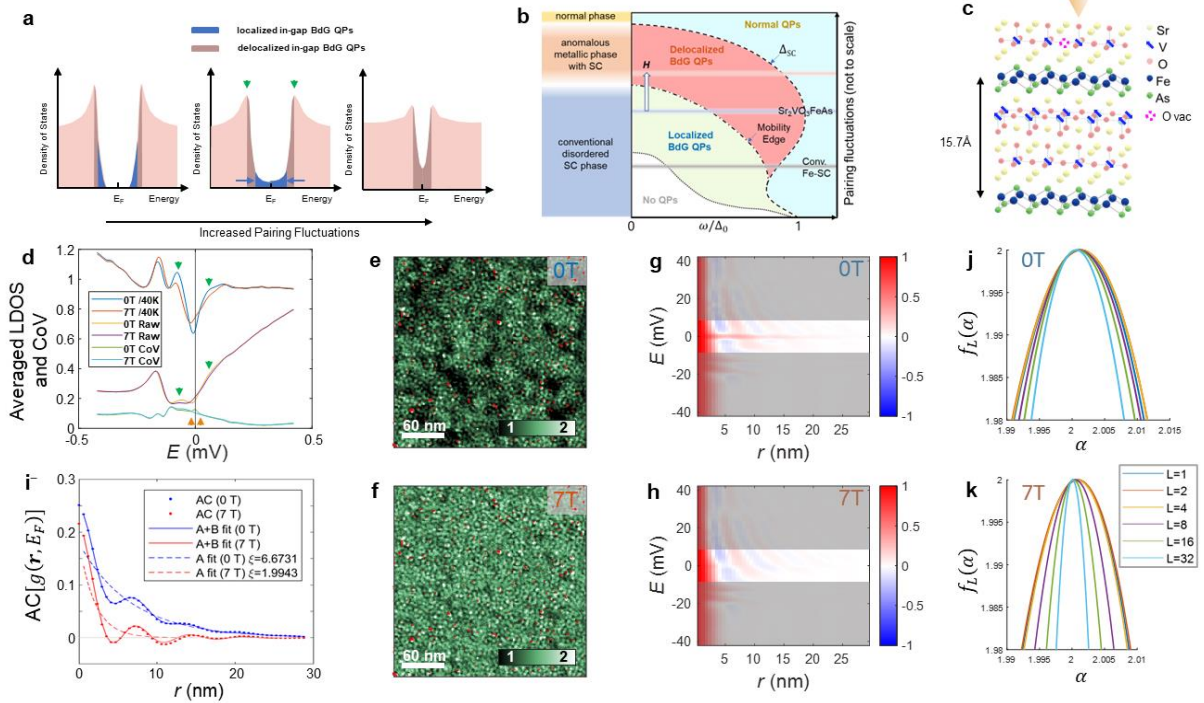

**Figure 1 | Schematic phase diagram, atomic structure, field-dependent LDOS, and autocorrelation length scale of Anderson localization in a quasi-2d disordered superconductor with zero-bias quasiparticles. a,** Schematic quasiparticle LDOS plots for disordered quasi-2d superconductor as pairing fluctuations are increased. The rightmost one shows an anomalous metallic phase with superconductivity at large pairing fluctuations. **b,** Schematic phase diagram for Anderson (de)localization of the in-gap BdG quasiparticles of quasi-two dimensional disordered superconductors in terms of energy and pairing fluctuations, based on this work. The lower part is based on our RG analysis (Fig. 4c) while the upper half is based on our experimental observation. **c,** Side view of charge-neutrally cleaved $Sr_2VO_3FeAs$ with a large inter-FeAs-layer spacing of 15.7Å and the Néel-ordered magnetic V atoms in the gapped $Sr_2VO_3$ layers padding the FeAs layers. The O vacancy defects play the role of the dominant potential scattering centers. **d,** Field-of-view-averaged tunneling spectra taken at 0 T and 7 T (middle), the same spectra divided by the reference spectrum taken at 0 T and 40 K (top), and their coefficients of variation (bottom). The coherence peaks and the mobility edges are marked with green and blue arrows respectively. **e-f,** Zero-bias LDOS maps at 0 T and 7 T with the $VO_2$-layer O-vacancy defects shown as overlaid red dots (See Ext. Data Fig. 2 for more details). As the 7 T external magnetic field is applied, the local density of states increases to become more homogeneous. **g-h,** Energy-resolved angle-averaged auto-correlations of LDOS maps at 0 T (**g**) and 7 T (**h**). Note the quasi-divergent length scale for density fluctuations, given by a sharp red signature near the zero energy, disappears by the applied magnetic field. **i,** The zero-bias cross-section data of **g** and **h** with normalization turned off. The experimental data (scattered dots) are fitted (solid lines) to the model $Ae^{-r/\xi} + Be^{-\kappa r}\cos(2k_F r - \phi)$ with the first term due to the LDOS modulation induced by the Anderson localization and the second term due to the quasiparticle interference. The dashed lines are the first fitted term $Ae^{-r/\xi}$. Long-range density fluctuations are suppressed by the applied magnetic field ($\xi_{0T} = 6.67$ nm → $\xi_{7T} = 1.99$ nm). See SI section 1.1 for the details of the fitted function and parameters. **j-k,** Scale-dependent multifractal spectra $f_L(\alpha)$ for LDOS at $E_F$ at 0 T (**e**) and 7 T (**f**) with measure size $L$ varied over 1, 2, 4, 8, 16 and 32 pixels with

conversion factor of 0.586 nm per pixel. The scale invariance terminates at the localization length scale. See the text, the SI section 1.3 and the Ext. Data Fig. 6 for more details.

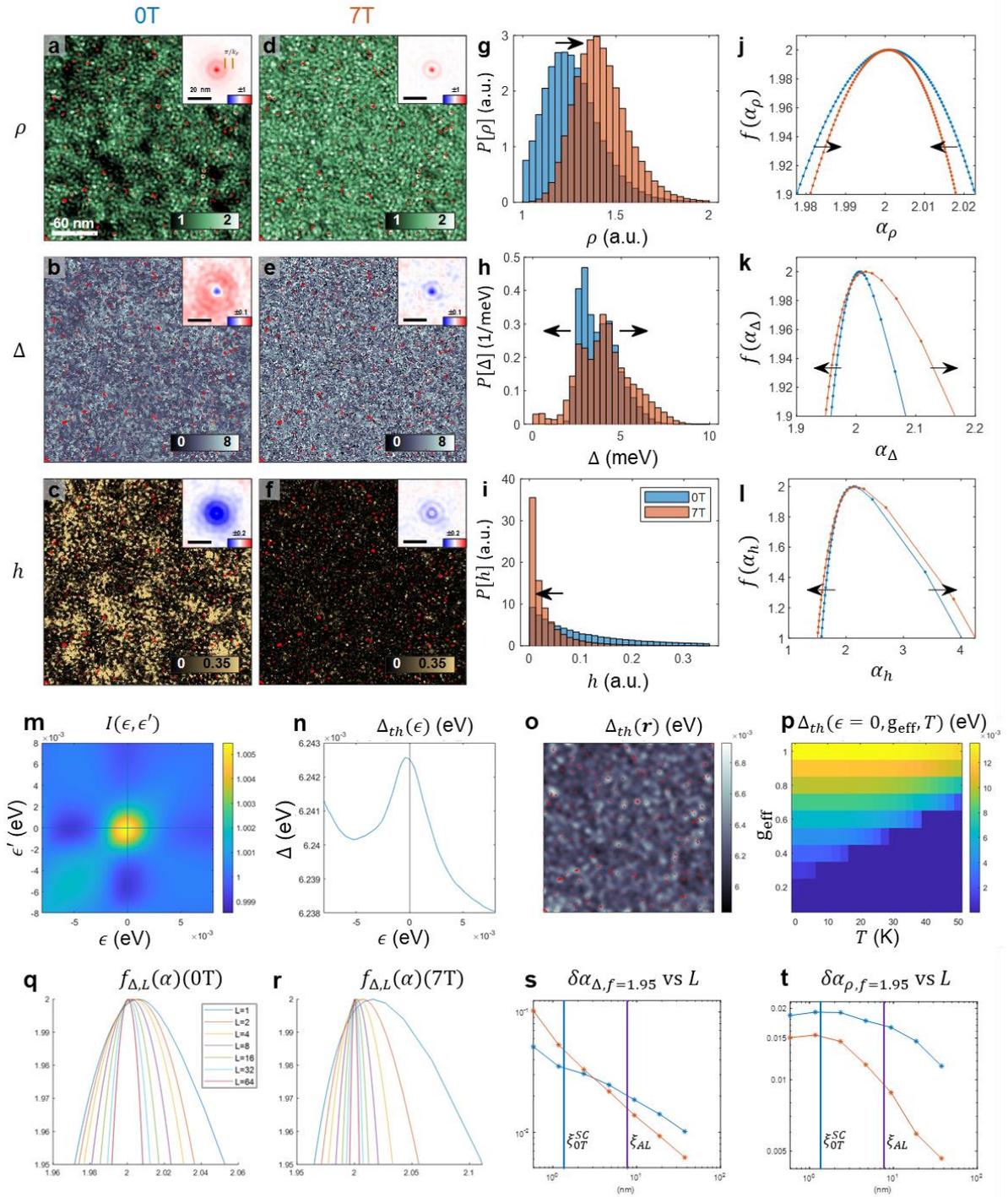

**Figure 2 | The experimental distribution functions and multifractal spectra of the zero-bias LDOS, the SC gap and the coherence peak height, with and without applied magnetic field. a-c,** Zero-bias LDOS map (**a**), gap map (**b**) and coherence peak height map (**c**) at 0 T. Inset of **a** is the auto-correlation map of **a** with the QPI modulation wavelength marked. **d-f,** Same as **a-c**, at 7 T. Insets of **b-f** are the cross-correlation maps of **a** and **b-f**. The overlaid red dots are the VO$_2$-layer O-vacancy defects. Note the zero-bias LDOS negatively correlated with both local pairing amplitude and coherence peak height. **g-i,** Histograms of the maps **a-f**. Note the distribution of the zero-bias LDOS is shifted to a larger average value with narrower width with applied magnetic field. The peak

of the local pairing gap (**h**) around 3 meV shows significant suppression with applied magnetic field and are transferred to those around the zero energy. The width of distribution of the local pairing gap is also enhanced with applied magnetic field. The distribution function of the coherence peak height (**i**) is suppressed with applied magnetic field. **j-l,** Multifractal spectra $f(\alpha)$ of the maps **a-f**. Note the multifractal spectrum of the local density of states gets narrower with applied magnetic field. On the other hand, both of the multifractal spectra of the local pairing gap and the coherence peak height get broader with applied magnetic field. **m-p**, To understand the origin of the positive long-range correlation between the zero-bias LDOS map and the local pairing-gap map, we solve the self-consistent gap equation in strongly disordered superconductors [18,21]. See the text for more details. **m**, $I(\epsilon,\epsilon') = \int d^d\vec{r} \, |\psi(\epsilon,\vec{r})|^2 \, |\psi(\epsilon',\vec{r})|^2$ is the autocorrelation function, where $|\psi(\epsilon,\vec{r})|^2$ is replaced with the LDOS measured in the superconducting state. The bright spot in $(\epsilon,\epsilon') = (0,0)$ indicates Anderson localization at zero magnetic field. **n**, Resorting to $I(\epsilon,\epsilon')$, we find the energy-dependent gap function $\Delta(\epsilon)$. **o**, We convert $\Delta(\epsilon)$ to its real space configuration $\Delta(\vec{r})$. Comparing Fig. 2o with Fig. 2a (Fig. 2b), we observe clear signature of the positive long-range correlation, which confirms that Anderson localization is responsible for this phenomenon. **p**, Repeating this analysis at finite temperatures, we obtain critical temperatures as a function of the BCS coupling constant, which gives reasonable values ~ 30K. **q-r**, Scale-dependent multifractal spectra $f_{\Delta,L}(\alpha)$ for SC gap map at 0 T(**m**) and 7 T(**n**) with measure size $L$ varied with conversion factor of 0.586 nm per pixel. **s-t**, The widths of the multifractal spectra (**o**) $f_{\Delta,L}(\alpha)$ and (**p**) $f_{\rho,L}(\alpha)$ measured at $f(\alpha_\pm) = 1.95$ as $L$ (horizontal axis) and magnetic field are varied. The width $\delta\alpha_\Delta(\alpha)$ at 0 T and 7 T show cross-over at a length scale in between SC coherence length ($\xi_{0T}^{SC}$) and AL correlation length ($\xi_{AL}$). The width $\delta\alpha_\rho(\alpha)$ shows scale-invariance for $L \lesssim \xi_{AL}$ at 0 T expected due to self-similarity of localized electronic wavefunctions.

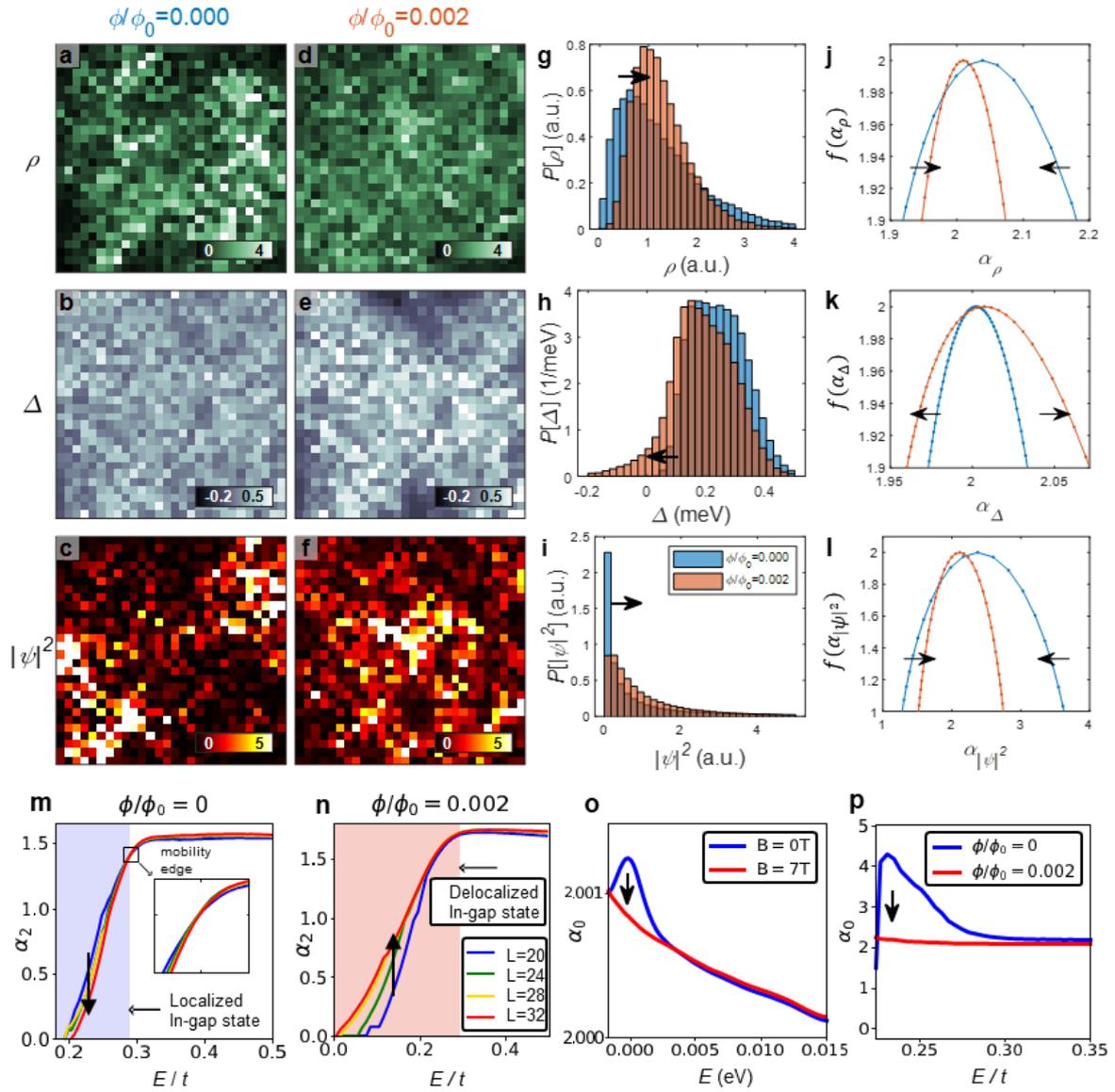

**Figure 3 | Hartree-Fock-BCS-Anderson simulated distribution functions and multifractal spectra of the low-bias LDOS, the SC gap and the wavefunction, with and without applied magnetic field. a,** Low-bias LDOS inside the superconducting gap (the energy window $E/t \sim [0.25, 0.27]$) at $\phi/\phi_0 = 0$ (zero-field condition). **b,** Gap map at $\phi/\phi_0 = 0$ for a specified disorder configuration. **c,** The probability density for $E/t = 0.26$ at $\phi/\phi_0 = 0$. **d,** Same as **a**, at $\phi/\phi_0 = 0.002$ (field condition). Note the LDOS is enhanced to be more homogeneous with applied magnetic field. **e,** Same as **b**, at $\phi/\phi_0 = 0.002$. Note the region of vanishing local pairing amplitudes is enlarged with applied magnetic field. **f,** Same as **c**, at $\phi/\phi_0 = 0.002$. Note the probability density of the in-gap quasiparticles gets uniform with applied magnetic field. **g-i,** Histograms of the maps **a-f**. Note the distribution function of the local pairing amplitude is broadened with applied magnetic field. In particular, there appears finite population of the vanishing local pairing amplitude and even negative sign of the local pairing gap is observed. On the other hand, the distribution function of the LDOS is narrowed and shifted to a larger average value. The distribution of the probability density for $E/t = 0.26$ is consistent with that of the LDOS. **j-l,** Multifractal spectra $f(\alpha)$ of the maps **a-f**. Note the multifractal spectrum of the local pairing gap is broadened with applied magnetic

field. On the other hand, both of the multifractal spectra of the LDOS and the probability density are narrowed with applied magnetic field. **m,** Multifractal critical exponent ($\alpha_2$) of the inverse participation ratio of the low-energy in-gap BdG quasiparticles in the Hartree-Fock-BCS-Anderson simulation. The scale-invariance near $E/t \sim 0.29$ identifies the mobility edge. **n,** Evolution of the mobility edge with applied magnetic field. Applied magnetic field turns an Anderson insulating state of the low-energy in-gap BdG quasiparticles into their metallic phase. There also exists a mobility edge at a high energy, not shown here and not affected by applied magnetic field. It is interesting to observe that BdG quasiparticles near the coherent peak become more critical with applied magnetic field. **o,** Experimental multifractal critical exponent ($\alpha_0$) for the LDOS of BdG quasiparticles as a function of energy with and without applied magnetic field, **p**, Same as **o**, for the simulated probability density. Note the applied magnetic field suppress the magnitude of $\alpha_0$ for low-energy in-gap BdG quasiparticles.

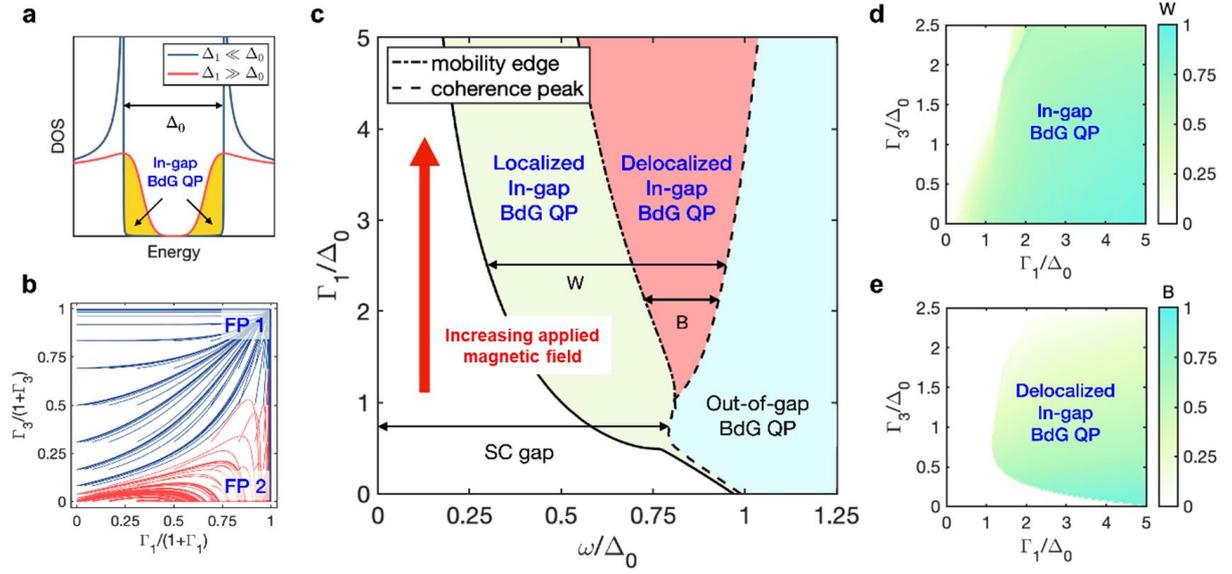

**Figure 4 | Renormalization group analysis with potential and pairing fluctuations showing the localization-delocalization transition of the in-gap BdG quasiparticles. a,** A schematic diagram of the density of states (DOS) showing the disorder-induced in-gap BdG quasiparticle states. Here, $\Delta_0$ and $\Delta_1$ stand for the pairing terms in the effective field theory in Eq. (1), which correspond to the energy gap and energy range of in-gap states, respectively. **b,** An example of the renormalization group flow for the effective field theory. Here, $\Gamma_1$ and $\Gamma_3$ are the disorder strengths for random superconducting order parameter fluctuations and random potential fluctuations for the in-gap BdG quasiparticles, respectively. At the stable fixed point of FP1 (FP2), $\Gamma_3$ grows (diminishes), indicating that the in-gap states are localized (delocalized). **c,** An energy-dependent phase diagram of BdG quasiparticle states as a function of quasiparticle energy $\omega$ and $\Gamma_1$. The delocalized in-gap BdG quasiparticle states (red region) emerge in disordered superconductors under external magnetic field ($\Gamma_1 \gg \Delta_0$). As $\Gamma_1$ increases, the minimum energy level for the delocalized states is shifted toward the Fermi level ($\omega = 0$), indicating that the delocalization phase transition occurs by the application of magnetic field which turns on the superconducting order parameter fluctuations ($\Gamma_1$). **d-e,** Phase diagrams showing the energy ranges of the full in-gap states ($W$) and delocalized in-gap states ($B$), respectively, which are drawn in **c**, as a function of $\Gamma_1$ and $\Gamma_3$.